\documentclass{svproc}
\usepackage{url}

\usepackage{amsmath,amssymb}
\usepackage{booktabs}
\usepackage{graphicx}
\usepackage{CJKutf8}
\usepackage{xcolor}
\usepackage[hidelinks]{hyperref}

\newcommand{\BRec}{\operatorname{BRec}}
\newcommand{\Rec}{\operatorname{Rec}}
\newcommand{\SC}{\operatorname{Capture}}
\newcommand{\Reward}{\operatorname{NetCapture}}

\begin{document}
\mainmatter
\title{Teaching Language Models to Narrow Boolean Patent Searches from Search Feedback}
\titlerunning{Teaching Language Models to Narrow Patent Searches}
\author{Seiya Kawano\inst{1,2} \and Koji Marusaki\inst{3} \and Hirofumi Nonaka\inst{4} \and Koichiro Yoshino\inst{5,2}}
\authorrunning{S. Kawano et al.}
\tocauthor{Seiya Kawano, Koji Marusaki, Hirofumi Nonaka, Koichiro Yoshino}
\institute{Kyoto Institute of Technology, Kyoto, Japan
\and Guardian Robot Project, RIKEN, Kyoto, Japan
\and Nagaoka University of Technology, Nagaoka, Japan
\and Aichi Institute of Technology, Nagoya, Japan
\and The University of Osaka, Osaka, Japan\\[6pt]
\email{kawano@kit.ac.jp}, \email{hnonaka@aitech.ac.jp}, \email{yoshino.koichiro.es@osaka-u.ac.jp}}

\maketitle
\setcounter{footnote}{0}

\begin{abstract}
Patent prior-art searchers issue Boolean queries line by line, broadening to find relevant documents and narrowing until candidates can be screened. We distill 38,397 Japan Patent Office search reports into executable expert strategies, re-execute them, and evaluate generated strategies by citation recall and the smallest set containing each citation. Prompted models write valid, high-recall strategies but leave citations in sets too large to screen. We adapt a generator offline from scored rollouts in two forms: strategy-level learning from a screening-aware strategy score, and line-level learning from citation-preserving narrowing credit, which upweights the query lines keeping a found citation in a smaller set. On shared rollouts, line-level credit outperforms strategy-level scores on screening measures; iterated, it cuts the median citation-containing set from 768 to under 50 and places more citations in sets under 100 documents than experts, at lower recall. Strategy-level preference learning reaches near-complete recall; the two forms are complementary.
\end{abstract}

\section{Introduction}\label{sec:intro}
Prior-art search decides whether an invention deserves a patent, and the same retrieval problem returns with higher stakes in invalidity and freedom-to-operate analysis. A disclosure missed at examination resurfaces years later in an invalidation trial, at far greater public cost. Searchers therefore cannot stop at a convenient cutoff, and the coverage they pursue is paid for in documents that somebody has to read~\cite{lupu2013patent}.

Professional practice differs from the ranked retrieval of web search. Searchers write Boolean queries that combine full text with patent classifications such as IPC, CPC, and the Japanese FI and F-terms. A Boolean strategy is auditable: the searcher can justify every retrieved document, record the searched scope officially, and trade recall against the number of documents to be screened. Where a retrieval decision can carry legal weight, this auditability is a requirement rather than a preference. Neural retrievers and LLM-based research agents are hard to audit document by document, and ranked screening with statistical stopping rules certifies coverage probabilistically rather than through a recorded query; these approaches address complementary problems (Sect.~\ref{sec:related}). Shared tasks in TREC, CLEF-IP, and NTCIR have shaped patent retrieval~\cite{lupu2009trec,piroi2011clef,fujii2007ntcir}, but they frame it as one-shot ranking. Real prior-art search is iterative: the searcher issues a query, inspects its hit count, broadens to discover new aspects, and narrows to isolate screenable sets. Success means retrieving the relevant documents within a set small enough to read, and recall alone cannot distinguish a citation buried among tens of thousands of documents from one delivered among a few dozen. Recent work optimizes query generation for retrieval metrics~\cite{nogueira2017query,jiang2025deepretrieval,wang2026autobool}, but mostly for a single query rather than a sequence that refines its own earlier results.

Expert search strategies are, however, recorded at scale. Patent offices publish search reports that document the strategy employed and the prior art cited. We treat these reports as supervision for the search process itself. We distill Japan Patent Office (JPO) records into 38,397 reports with 425,405 executable Boolean query lines linked to 109,717 cited prior-art occurrences, and build a search environment that re-executes expert and generated strategies while recording, for every line, its exact hit count and whether each target citation is present.

\begin{figure}[t]
\centering
\includegraphics[width=\textwidth]{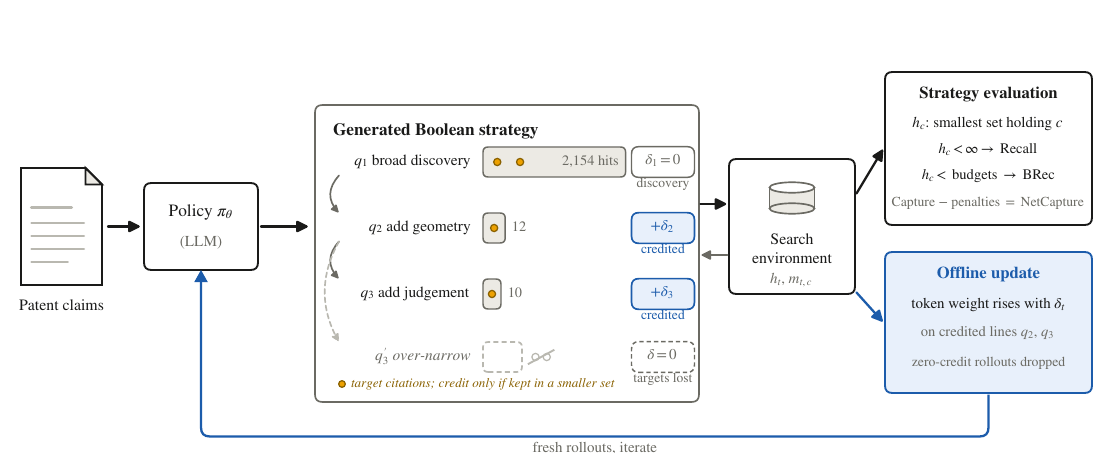}
\caption{Overview of the line-level form (narrowing credit) with a schematic strategy. The environment executes each generated line and observes its hit count $h_t$ and citation membership $m_{t,c}$ (gold dots mark the two target citations). Evaluation takes, per citation, the smallest set that contains it. A line earns credit $\delta_t$ only by keeping an already-found citation in a smaller set ($q_2$, $q_3$); first discovery earns none ($q_1$), and a hypothetical over-narrowing alternative that loses the citations earns nothing ($q_3'$).}
\label{fig:overview}
\end{figure}

On this resource we adapt a language-model generator offline from search feedback in two forms. Strategy-level learning attaches our score, NetCapture, to each strategy and trains with existing learners. A complete-strategy score indicates which strategy is better, but not which line performed the useful refinement, so we also propose line-level learning from \emph{citation-preserving narrowing credit}: a later query line receives token-level training weight when it retrieves a citation already found by an earlier line while reducing the size of the containing set. Broad discovery is allowed and not penalized. Figure~\ref{fig:overview} summarizes the task and the line-level training signal. Our contributions are as follows.
\begin{itemize}
\item We build a dataset of executable expert search strategies linked to their cited prior art,\footnote{The dataset is available at \url{https://doi.org/10.6084/m9.figshare.33325407}.} together with a search environment that re-executes them (Sect.~\ref{sec:data}).
\item We define a task and an evaluation that measure, in addition to recall, the smallest set in which each citation is reached, summarized by a screening-aware score, NetCapture, and show that prompted language models find citations but do not narrow them into screenable sets (Sects.~\ref{sec:evaluation} and~\ref{sec:results-prompt}).
\item We adapt the generator from search feedback in two forms: strategy-level learning from NetCapture with existing learners, and line-level learning from citation-preserving narrowing credit, which we propose. Controlled comparisons on shared rollouts show that weighting the narrowing lines outperforms strategy-level scores and uniform imitation of the same rollouts, and under iteration the two forms reach different operating points between coverage and screening (Sects.~\ref{sec:adaptation} and~\ref{sec:results}).
\end{itemize}

\section{Related Work}\label{sec:related}
Patent retrieval has been studied through shared tasks~\cite{lupu2009trec,piroi2011clef,fujii2007ntcir} and surveyed broadly~\cite{shalaby2019patent,krestel2021survey}, as have neighboring patent tasks such as analysis~\cite{marusaki2024study} and drafting~\cite{kawano2024claimbrush,le2026large}. Work on query formulation extracts effective queries from application text~\cite{xue2009patent} and expands them with citations or term selection~\cite{mahdabi2014citation,golestanfar2015term,marusaki2026synonym}; evaluation in all of these scores one ranked list per topic. Technology-assisted review ranks a given candidate pool with active learning~\cite{cormack2014tar} and stops by statistical rules~\cite{lewis2021certifying}. It optimizes the reading of a candidate set, whereas we address the construction of that set.

Learning query reformulation from retrieval feedback goes back to reinforcement learning for ad-hoc search~\cite{nogueira2017query}, and recent systems couple generation with live engines~\cite{jin2025searchr1,jiang2025deepretrieval}. Boolean query generation has been studied most for systematic reviews~\cite{scells2020boolean,wang2023chatgpt}; closest to us, AutoBool trains a model with group relative policy optimization (GRPO) to emit one Boolean query per topic, rewarding both recall and a recall-modulated precision term~\cite{wang2026autobool}. A patent strategy is instead a sequence of lines that refer to and subtract earlier results, and uncited documents are unlabeled rather than irrelevant, so precision of one result set cannot be the target.

Our strategy-level conditions attach one scalar outcome to a complete response, either as reward-weighted regression~\cite{peng2019awr,peters2007rwr,eysenbach2023onestep} or as preference pairs~\cite{rafailov2023dpo}. The proposed credit is a process signal in the sense of process supervision~\cite{uesato2022process,lightman2024verify}, computed from search observations rather than from step annotations, and the iterative loop resembles Reinforced Self-Training~\cite{gulcehre2023rest} with a fixed search environment in place of a learned reward model.

\section{Dataset of Expert Search Strategies}\label{sec:data}
In Japan, part of the examiner's prior-art search is commissioned to registered search organizations, and the resulting search report becomes a public patent document once examination begins. A report contains a query table, with numbered Boolean lines and the hit counts observed by the searcher, and a cited-document table that links prior-art publications to the lines that retrieved them. Lines combine Boolean union (\texttt{+}), intersection (\texttt{*}), and difference (\texttt{-}), wildcards, proximity windows, field restrictions such as full text (\texttt{/TX}), classifications (IPC, FI, F-terms), and references to earlier result sets (\texttt{\char`\\$k$}).

We retain reports whose queries execute in a standard Boolean search system and whose citations resolve to Japanese publications in our collection: 38,397 of 106,550 scanned reports, excluding mainly foreign-theme reports and unparseable queries. Table~\ref{tab:data} gives the composition. Splits are disjoint by application. Since 99.3\% of test themes also occur in training, generalization is within familiar technical areas.

\begin{table}[t]
\centering
\caption{Search-report dataset composition.}
\label{tab:data}
\small
\setlength{\tabcolsep}{8pt}
\begin{tabular}{lrrr}
\toprule
Statistic & Train & Development & Test \\
\midrule
Reports & 37,668 & 373 & 356 \\
Query lines & 417,215 & 4,110 & 4,080 \\
Lines/report, median & 10 & 9 & 10 \\
Lines/report, mean & 11.08 & 11.02 & 11.46 \\
Citation occurrences & 107,549 & 1,124 & 1,044 \\
Search themes & 2,708 & 572 & 548 \\
\bottomrule
\end{tabular}
\end{table}

A report is a small Boolean program rather than a bag of queries. In the test split, 67.6\% of lines refer to an earlier result set and 65.4\% subtract one, while 71.9\% use a text field and 38.5\% a patent classification. In Fig.~\ref{fig:example-report}, an F-term intersection sets the topic, an IPC restriction isolates a sub-question, and text refinements subtract what has been inspected. We re-execute these strategies in our environment and use them as the \emph{expert reference}. It is a practice reference rather than an upper bound, since the searcher worked interactively and also pursued uncited relevant art; hit counts printed in the reports are not used in evaluation.

\begin{figure}[t]
\centering
\begin{CJK}{UTF8}{ipxm}
\footnotesize
\begin{tabular}{@{}c l r@{}}
\toprule
No. & Query line & Hits \\
\midrule
1 & \texttt{AA02*AA06*BA01*FA32} & 147 \\
2 & \texttt{H01Q(21/08*15/16)-\char`\\1} & 7 \\
3 & \texttt{AA02*BA01*水平偏波/TX-\char`\\1-\char`\\2} & 76 \\
4 & \texttt{AA02*(反射*水平偏波)/TX-\char`\\1-\char`\\2-\char`\\3} & 120 \\
5 & \texttt{(反射*水平偏波*合成)/TX-\char`\\1-\char`\\2-\char`\\3-\char`\\4} & 195 \\
\bottomrule
\end{tabular}
\end{CJK}
\caption{An expert strategy from a test report on an antenna theme. \texttt{AA02} etc.\ are F-terms, \texttt{H01Q(\ldots)} is an IPC restriction, and \texttt{-\char`\\$k$} subtracts the result of line $k$; the Japanese terms mean horizontal polarization, reflection, and combining. The two citations are associated with lines 2 and 3. Hit counts are as recorded in the report.}
\label{fig:example-report}
\end{figure}

\section{Sequential Boolean Query Generation}\label{sec:evaluation}
The task follows the trial-and-error procedure of a prior-art searcher. Given an application, the searcher must find every relevant prior-art document, and must do so within sets small enough to be read in full. Each query line is one search: it returns a set of documents, the searcher sees how many there are, and the next line is written in response, adding a term to narrow the previous one, intersecting or subtracting earlier results, or opening a new direction. The record of such a session is a sequence of query lines, and that sequence is what we generate. A strategy succeeds for a relevant document when some line returns it in a small set, and fails when no line returns it; a line that returns it among thousands of documents does not help. In evaluation, the relevant documents are the citations recorded for the application, from a few to a few dozen.

Given the claim text of an application and candidate theme codes taken from its report, which scope the search by JPO F-term technical areas, a model writes query lines $q_1,\ldots,q_T$ in the grammar above; the complete strategy is generated before execution, and search feedback is used only to train the generator (Sect.~\ref{sec:adaptation}). The search environment indexes unexamined and granted Japanese patent publications with character n-grams, expands classification codes under the same fixed rules for all conditions, and excludes documents published after each report's search date.

For a report, let $C$ be the set of target citations. Line $q_t$ retrieves $h_t$ documents, and $m_{t,c}=1$ if citation $c$ is among them. The smallest set reaching $c$ is $h_c=\min\{h_t:m_{t,c}=1\}$, with $h_c=\infty$ if no line retrieves it. Recall is $\Rec=|C|^{-1}\sum_{c}\mathbf{1}[h_c<\infty]$. For screening budgets $\mathcal{K}=\{50,100,200,500,1000,2000\}$, strict recall and Budgeted Recall are
\begin{equation}
\Rec_{<k}=\frac{1}{|C|}\sum_{c\in C}\mathbf{1}[h_c<k],\qquad
\BRec=\frac{1}{|\mathcal{K}|}\sum_{k\in\mathcal{K}}\Rec_{<k}.
\end{equation}
Min-hits reports the median and mean of $h_c$ over retrieved citations. Because $\Rec_{<k}$ changes only when a set crosses a budget, we also define a smooth score that gives partial credit above the budget, $g_k(h)=1$ for $h\le k$, $(k/h)^{1/2}$ for $k<h<\infty$, and $0$ for $h=\infty$ (the two conventions differ only at $h=k$). With $\SC$ the mean of $g_k(h_c)$ over citations and budgets, the summary score is
\begin{equation}
\Reward=\SC-0.01\max(0,T-10)-0.1N_{\mathrm{invalid}},
\end{equation}
where ten lines is the expert median and $N_{\mathrm{invalid}}$ counts lines that do not execute. Uncited documents are unlabeled rather than negative, which is why we measure set size instead of precision: set size counts how many documents surround a known citation without labeling any of them. Tables pool the recall measures and Min-hits over all test citations and average NetCapture over reports; validity (Val.\%) is the share of executable lines. These measures pick out the aspects that practitioners weigh when they judge a strategy: coverage, the size of the sets to be screened, length, and validity. The set-size measures ($\Rec_{<k}$, $\BRec$, Min-hits, and $\SC$) are per-citation proxies for screenability: each takes the best set in which a citation can be read, not the total reading of a session (Sect.~\ref{sec:limitations}).

\section{Citation-Preserving Narrowing Credit}\label{sec:adaptation}
We adapt the generator in three decoupled stages. Supervised fine-tuning (SFT) on report-derived strategies teaches the format and grammar. The policy then samples several strategies per application across a block of training reports, and the environment executes each line to return $h_t$ and $m_{t,c}$. Finally, these observations are converted into token weights for one epoch of weighted cross-entropy. The one-step setting applies this once to rollouts from the SFT policy; the iterative setting repeats it with fresh rollouts from the current policy on disjoint application blocks.

Complete-strategy scores do not tell which line made a citation screenable, so we build the signal per line. With the screening value $s(h)=1/(1+(h/100)^2)$, let $E_{t,c}=\{j<t:m_{j,c}=1\}$ be the earlier lines that retrieved $c$ and $b_{t-1,c}=\max_{j\in E_{t,c}}s(h_j)$ their best value. The credit of line $t$ is
\begin{equation}\label{eq:delta}
\delta_t=\frac{1}{|C|}\sum_{c\in C}\mathbf{1}[m_{t,c}=1\land E_{t,c}\neq\emptyset]\,\max\bigl(0,s(h_t)-b_{t-1,c}\bigr).
\end{equation}
The first retrieval of a citation sets the baseline but earns no credit, so broad queries are allowed, and a later line is rewarded only when it keeps the citation and improves its screening value. Splitting a narrowing into small steps does not increase the total credit, which depends only on a citation's screening value at first discovery and the best value later reached. Rollouts with $\sum_t\delta_t=0$ are excluded.

The credit becomes line weights
\begin{equation}
v_t=\epsilon+T(1-\epsilon)\frac{\delta_t}{\sum_j\delta_j},\qquad\epsilon=0.01,
\end{equation}
which have mean one within a strategy; a line without credit keeps the context weight $\epsilon$. Tokens of line $t$ receive $v_t$, the end marker receives $\epsilon$, and the loss is the weighted cross-entropy averaged over response tokens only. The loss contains no negative signal: a line that loses citations or returns nothing is treated as context rather than punished.

\section{Experimental Settings}\label{sec:settings}
\paragraph{Compared methods.} All adapted conditions start from the same SFT policy of Qwen3.5-9B,\footnote{\url{https://huggingface.co/Qwen/Qwen3.5-9B}} trained with LoRA~\cite{hu2022lora} on a broad report archive (excluding all development and test applications) and then on the training split. Prompt-only conditions use GPT-5.4-mini and the raw Qwen3.5-9B and Qwen3.6-35B\footnote{\url{https://huggingface.co/nvidia/Qwen3.6-35B-A3B-NVFP4}} with zero-shot (ZS) and few-shot (FS) prompts, and a text-field-only zero-shot prompt (TX ZS). All adapted conditions learn from the same search observations and differ only in how the observations become a training signal. The strategy-level conditions assign each strategy a single score (Recall, Budgeted Recall, or our NetCapture of Sect.~\ref{sec:evaluation}), so DPO and AWR on NetCapture are the strategy-level form of our search-feedback adaptation rather than external baselines. \emph{Advantage-weighted regression} (AWR)-style weighting standardizes the score within the rollouts of one application and weights cross-entropy by its exponential~\cite{peng2019awr}, and \emph{direct preference optimization} (DPO)~\cite{rafailov2023dpo} learns from pairs ordered by the score. \emph{Credit-pairing} DPO pairs rollouts of equal recall and prefers the one with the larger total credit $\sum_t\delta_t$, and \emph{filtered behavior cloning} (Filtered BC) isolates the line weighting: it trains on the same rollouts as the proposed update, those with at least one narrowing event ($\sum_t\delta_t>0$), but with plain response-only cross-entropy, giving every line the same weight.

\paragraph{Rollouts and training.} Each update samples eight rollouts for each of 256 training reports at temperature 0.9, top-$p$ 0.95, top-$k$ 50, and repetition penalty 1.08, with 5 to 20 lines per strategy. A time cap on search, applied without regard to the scores, leaves 2 to 8 scored rollouts per report (1,712 to 1,832 per update). All one-step conditions share the same scored rollouts. Iterative adaptation runs for the proposed credit (five updates), for AWR with Budgeted Recall (stopped at three, its development recall never reaching the 0.90 floor) and NetCapture (five), and for DPO with NetCapture (five). For the proposed credit the reported update is chosen on 50 development reports by a fixed rule: the highest development NetCapture among updates with recall at least 0.90 and validity at least 0.99; strategy-level conditions report their final update. LoRA (rank 8, scaling 16, dropout 0.05) is applied to all attention and MLP projections, with AdamW, learning rate $5\times10^{-5}$, and one epoch; DPO uses learning rate $10^{-5}$, $\beta=0.1$, and the policy at the start of each update as reference. The test set has 356 reports and 1,042 unique target citations (1,044 occurrences in Table~\ref{tab:data}), with one strategy generated per report under a fixed seed.

\section{Results}\label{sec:results}
\begin{table}[t]
\centering
\caption{Test results on 356 reports. Min-hits is lower-is-better; all other columns are higher-is-better. Bold marks the best value among generated strategies. NetCapture-based AWR and DPO and narrowing-credit learning are the strategy-level and line-level forms of our search-feedback adaptation.}
\label{tab:main}
\scriptsize
\setlength{\tabcolsep}{1.8pt}
\begin{tabular}{@{}lrrrrrrrr@{}}
\toprule
Method & NetCap. & Recall & BRec & Rec$_{<100}$ & Rec$_{<500}$ & Rec$_{<1000}$ & \shortstack[r]{Min-hits\\med./mean} & Val.\% \\
\midrule
Expert reference & .6997 & .9655 & .4938 & .2821 & .6142 & .7179 & 282/1626.3 & 99.85 \\
\midrule
GPT-5.4-mini TX ZS & .5438 & .8954 & .3330 & .1363 & .3896 & .5182 & 666/2795.2 & 98.20 \\
GPT-5.4-mini ZS & .5504 & .8196 & .3354 & .1603 & .3944 & .5106 & 541/2883.6 & 99.55 \\
GPT-5.4-mini FS & .5749 & .8800 & .3228 & .1286 & .3762 & .5115 & 681/3103.4 & 99.68 \\
Qwen3.5-9B ZS & .5304 & .9741 & .3023 & .1046 & .3551 & .4875 & 995/2614.2 & 100.00 \\
Qwen3.5-9B FS & .4154 & .7850 & .2815 & .1459 & .3042 & .4146 & 924/3010.3 & 99.99 \\
Qwen3.5-9B TX ZS & .5212 & .9434 & .3300 & .1363 & .3685 & .5115 & 843/2475.2 & 99.99 \\
Qwen3.6-35B ZS & .5203 & .8532 & .3954 & .2159 & .4463 & .5873 & 457/1967.0 & 100.00 \\
Qwen3.6-35B FS & .3134 & .5528 & .2585 & .1468 & .3090 & .3743 & 383/2081.3 & 100.00 \\
Qwen3.6-35B TX ZS & .5811 & .9626 & .3941 & .1737 & .4568 & .5998 & 589/1845.8 & 99.99 \\
\midrule
\multicolumn{9}{@{}l}{\emph{One-step adaptation from the shared scored rollouts}}\\
AWR Recall-only & .4968 & .8944 & .3519 & .1564 & .4155 & .5441 & 634.5/2016.1 & 99.68 \\
AWR Budgeted-Recall & .5176 & .8916 & .3730 & .1737 & .4223 & .5835 & 541/1724.4 & 99.30 \\
DPO Recall-only & .4943 & .9271 & .3385 & .1488 & .3618 & .5154 & 823/1964.0 & 99.93 \\
DPO Budgeted-Recall & .5077 & .8954 & .3452 & .1574 & .3781 & .5298 & 687/1687.1 & 99.93 \\
DPO Credit pairing & .4530 & .8292 & .3145 & .1478 & .3369 & .4808 & 691/1955.7 & 99.83 \\
Filtered BC & .5143 & .8848 & .3560 & .1593 & .4079 & .5355 & 602.5/1684.9 & 99.27 \\
Ours: NetCapture-AWR & .5422 & .8829 & .3888 & .1910 & .4386 & .5960 & 502.5/1637.5 & 99.97 \\
Ours: NetCapture-DPO & .5037 & .9155 & .3389 & .1382 & .3599 & .5384 & 737/1954.6 & 99.97 \\
Ours: Narrowing credit & .5952 & .9155 & .4403 & .2322 & .5029 & .6363 & 393/1495.9 & 99.89 \\
\midrule
\multicolumn{9}{@{}l}{\emph{Iterative adaptation (fresh on-policy rollouts at every update)}}\\
SFT (initialization) & .4346 & .7975 & .3017 & .1401 & .3349 & .4607 & 768/2340.5 & 100.00 \\
AWR Budgeted-Recall & .5313 & .8656 & .4071 & .2073 & .4808 & .5931 & 360.5/1443.1 & 99.97 \\
Ours: NetCapture-AWR & .3426 & .6881 & .2817 & .1440 & .3138 & .4261 & 645/2216.1 & 98.72 \\
Ours: NetCapture-DPO & \textbf{.7493} & \textbf{.9923} & \textbf{.6655} & .4434 & \textbf{.7716} & \textbf{.8829} & 129.5/551.6 & 100.00 \\
Ours: Narrowing credit & .6774 & .8752 & .6636 & \textbf{.5509} & .7294 & .7898 & \textbf{48.5}/\textbf{448.6} & 100.00 \\
\bottomrule
\end{tabular}
\end{table}

\subsection{Prompting Alone}\label{sec:results-prompt}
Prompted models rarely break the grammar: validity is near 100\% for the open models, and raw Qwen3.5-9B zero-shot even exceeds the expert reference in recall (Table~\ref{tab:main}). The citations they retrieve, however, sit in large sets. For raw 9B zero-shot the median set containing a citation is more than three times the expert's, and every prompted condition is below the expert reference in $\Rec_{<100}$. Few-shot examples make recall drop sharply for both open models, suggesting that what transfers from demonstrations is the surface form of queries rather than the search procedure. Valid, high-recall strategies are therefore within reach of prompting; what is missing is the narrowing of found citations into sets that can be read.

\subsection{One-Step Adaptation}
All one-step conditions share the same scored rollouts. Strategy-level scores do improve on SFT, but what improves depends on the score: Recall-only DPO reaches the highest recall among them while half of its citations remain in sets above 800 documents, and switching to Budgeted Recall shrinks the sets at a cost in coverage. Even with NetCapture as the score, the best strategy-level conditions leave citations in sets of several hundred documents. The proposed credit exceeds every strategy-level condition in NetCapture, BRec, and Min-hits. Two controls locate the source of the difference. Filtered BC, which trains on the same rollouts with uniform line weights, recovers only about half of the NetCapture gain over SFT, so selecting rollouts is not enough and the weighting inside them carries the signal. Credit-pairing DPO, which projects the same credit onto strategy-level pairs, has lower recall than Recall-only DPO and no higher NetCapture, so the credit works as weights on the narrowing lines themselves, not as a preference between strategies.

\subsection{Iterative Adaptation}\label{sec:results-iter}
Repeating rollout generation, scoring, and the weighted update moves the policy well beyond the one-step result. The selected policy improves NetCapture, BRec, and recall over its SFT initialization and shrinks the median citation-containing set to about one sixteenth (from 768 to 48.5); NetCapture rises over SFT on 264 of the 356 test reports, 254 of them without recall loss. It trails the expert reference, and its own one-step version, in recall but reaches citations in far smaller sets, and its $\Rec_{<100}$ is well above every prompted condition and the expert reference.

\begin{figure}[t]
\centering
\includegraphics[width=\textwidth]{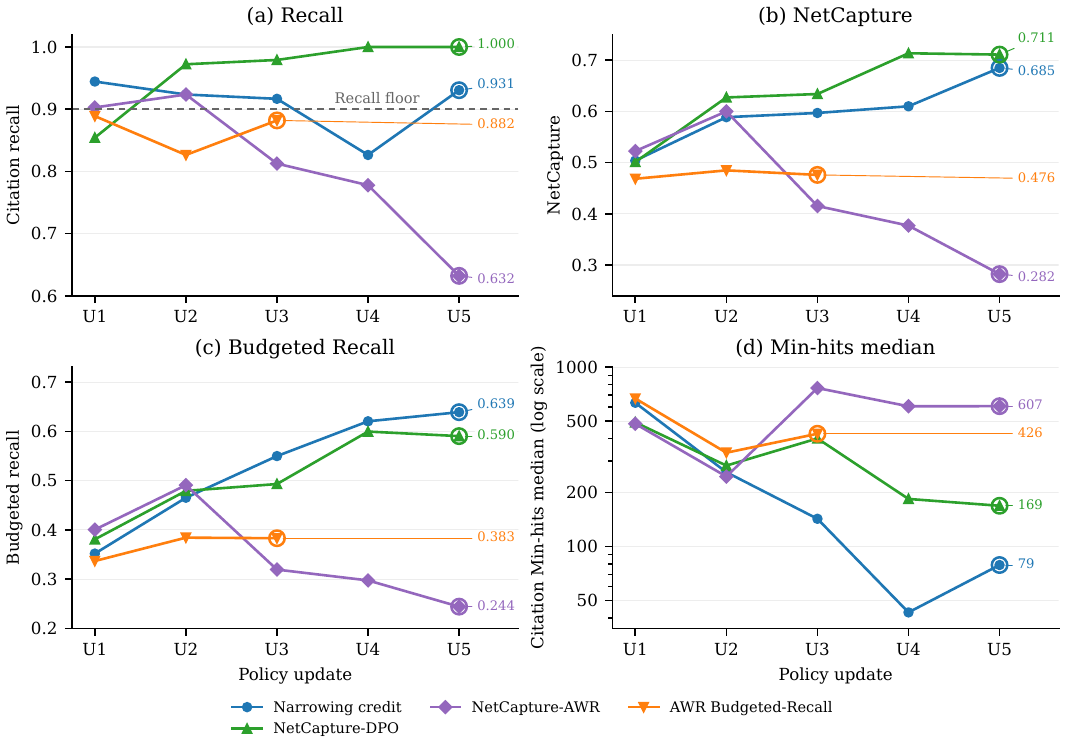}
\caption{Development trajectories of every iterative series (first 50 development reports): recall with the 0.90 floor, NetCapture, Budgeted Recall, and median Min-hits on a log scale. Rings mark the reported checkpoint of each series.}
\label{fig:iterative-curve}
\end{figure}

Iterating strategy-level scores gives results that depend on both the score and the learning rule. Three AWR updates with Budgeted Recall narrow sets somewhat over the one-step version with little NetCapture gain, and AWR with NetCapture degrades the policy (Fig.~\ref{fig:iterative-curve}). DPO with NetCapture, in contrast, works clearly: coverage becomes nearly complete and its NetCapture exceeds both the line-credit policy and the expert reference. Its gain, however, lies on the coverage side. Its BRec is level with the line-credit policy, its median citation-containing set is more than twice as large, and its $\Rec_{<100}$ is lower. Strategy-level NetCapture preferences and per-line credit thus reach different operating points, coverage-first and screening-first respectively. Under NetCapture a missed citation scores zero at every budget, and in DPO's trajectory, coverage indeed rises first and narrowing follows (Fig.~\ref{fig:iterative-curve}).

The development trajectories also show a cost of the per-line credit (Fig.~\ref{fig:iterative-curve}). As iteration deepens narrowing, coverage recedes: development recall falls below the 0.90 floor at update 4 before recovering at update 5, which the selection rule picks. The line that first retrieves a citation earns no credit and keeps only the context weight, so under iteration on its own rollouts the policy can lose the discovery behavior on which later narrowing depends. Because update 5 ends the series, whether further updates would sustain this recovery remains open.

\subsection{Learned Search Behavior}\label{sec:results-behavior}
Table~\ref{tab:behavior} compares behavior before and after adaptation for the narrowing-credit policy. Both SFT and the adapted policy run to the 20-line cap in most reports, so the gains do not come from issuing more queries. Narrowing events, lines with $\delta_t>0$, more than triple per report, and the share of reports with at least one event grows from under 40\% to over 80\%, while the number of lines that first retrieve a citation does not fall. The repertoire of query operations, however, contracts relative to expert practice. SFT tracks the expert's feature usage closely; the adapted policy narrows almost entirely through text, and classification and subtraction nearly disappear. The credit accrues wherever narrowing happens, text reformulation is the easiest way to narrow, and nothing in the objective distinguishes a classification restriction from an added text term. Subtracting an earlier set that contains a found citation cannot earn credit for that citation by construction, because the line no longer contains it; experts, however, subtract exactly what they have already inspected in order to reach new candidates (Fig.~\ref{fig:example-report}). Accordingly, the expert strategies show few narrowing events in this sense and the most discovery lines (Table~\ref{tab:behavior}): the credit teaches a different operation toward the same goal of a screenable set.

\begin{table}[!ht]
\centering
\caption{Search behavior on the test reports. The expert column is computed from the re-executed expert strategies with the same definitions.}
\label{tab:behavior}
\small
\begin{tabular}{lrrr}
\toprule
Measure & Expert ref. & SFT & Narrowing credit \\
\midrule
Narrowing events per report & 0.365 & 0.767 & 2.463 \\
Reports with $\ge$1 narrowing event & 21.6\% & 38.8\% & 82.3\% \\
New-citation discovery lines per report & 1.857 & 1.424 & 1.537 \\
\midrule
Lines using the TX text field & 71.9\% & 70.7\% & 86.3\% \\
Lines using a patent classification & 38.5\% & 36.2\% & 2.4\% \\
Lines referencing an earlier set & 67.6\% & 67.5\% & 12.6\% \\
Lines subtracting an earlier set & 65.4\% & 62.4\% & 4.4\% \\
Lines using proximity operators & 40.4\% & 39.2\% & 17.8\% \\
\bottomrule
\end{tabular}
\end{table}

\section{Limitations}\label{sec:limitations}
Our evaluation measures, for each citation, only the smallest set that reaches it. In practice searchers also read documents that recur across sets, so a sequence of nearly identical narrow text queries could reach a small set without reducing what a searcher reads; measuring effort by the union of returned sets is a natural next step. The narrowing-credit policy narrows mainly by text reformulation and rarely uses classifications or set algebra, which experts rely on to cover a field and justify its scope. Patent-specific knowledge, classification definitions, or credit for classification-based narrowing could restore that repertoire, and keeping the weight of the lines that discover citations could sustain coverage under iteration. Finally, our experiments cover Japanese reports in one query grammar, and strategies are generated in full before execution. Because the credit depends only on search observations, it applies wherever an executable search environment exists, including interactive line-by-line generation.

\section{Conclusion}\label{sec:conclusion}
We distilled JPO search reports into executable expert strategies, evaluated generated strategies by both recall and the smallest set containing each citation, and adapted a generator from search feedback in two forms: strategy-level learning with NetCapture, and line-level learning with citation-preserving narrowing credit, which weights the query lines that keep a found citation in a smaller set. Prompted language models find citations but leave them in sets too large to screen. On shared rollouts, localized credit outperforms strategy-level scores on the screening measures, and iterating it cuts the median citation-containing set by more than an order of magnitude, while preference learning on the strategy-level NetCapture score, the other form of our adaptation, reaches a coverage-first operating point; the two forms are complementary. Two challenges remain: preventing iterative narrowing from eroding coverage, and recovering the expert use of classifications and set algebra while keeping the screening gains.

\subsubsection*{Acknowledgments.}
This work was supported by the NEDO Wakasapo project (25101271-0), the Okawa Foundation Research Grant, and JST ACT-X (JPMJAX22A4).

\bibliographystyle{splncs03}
\bibliography{references}
\end{document}